# Simulation of Repair on Dynamic Patient-Specific Left Atrioventricular Valve Models

Stephen Ching BS[1], Christopher Zelonis MS[1], Christian Herz MS[1], Patricia Sabin MS[1], Matthew Daemer BS[1], Muhammad Nuri MD[2], Yan Wang RDCS[3], Devin W. Laurence PhD[3], Jonathan M. Chen MD[2], Lindsay S. Rogers MD[3], Michael D. Quartermain MD[3], John Moore MS[4], Terry Peters PhD[5], Elvis Chen PhD[5], Matthew A. Jolley MD[1,3]

[1]Children's Hospital of Philadelphia, Department of Anesthesia and Critical Care Medicine, Philadelphia, PA

[2]Children's Hospital of Philadelphia, Division of Cardiothoracic Surgery, Philadelphia, PA

[3]Children's Hospital of Philadelphia, Division of Cardiology, Philadelphia, PA

[4]Archetype Medical Inc., London, Ontario, Canada

[5]Robarts Research Institute, Western University, London, Ontario, Canada

Corresponding Author:

Matthew A. Jolley, MD

3401 Civic Center Blvd

Philadelphia, PA 19104

**Abstract**

*Purpose*: To develop and evaluate a dynamic, image-derived patient-specific physical simulation platform for the assessment of left atrioventricular valve (LAVV) repair strategies in pediatric patients with repaired atrioventricular canal defects.

*Methods:* 3D transesophageal echocardiographic images of two patients with regurgitant LAVVs were identified from an institutional database. Custom code in SlicerHeart was used to segment leaflets, define the annulus, and generate patient-specific valve molds. Silicone valve models were fabricated and tested in a pulse duplicator under simulated physiological conditions. Five unrepaired valves were analyzed for manufacturing consistency, and multiple surgical repair techniques were compared for two patient-specific models.

*Results*: Manufacturing variability was low in annular metrics (CV for annular circumference: 2.1%; commissural distance: 4.1%; annulus height: 14.7%) but higher in leaflet closure metrics (billow height: 11.1%; billow volume: 18.9%; tenting height: 45.9%; tenting volume: 73.5%). In Patient 1, cleft closure and an Alfieri stitch both eliminated the regurgitant orifice area, but the Alfieri stitch resulted in elevated mean pressure gradient (17 mmHg vs. 4-9 mmHg for other repairs) and deteriorated with repeated loading. In Patient 2, no repair eliminated regurgitation entirely; however, combining an 11 mm patch augmentation with commissuroplasty reduced regurgitant area to 0.147 $cm^2$, the smallest observed among tested strategies.

*Conclusion*: This study demonstrates the feasibility of a dynamic physical simulation platform for preclinical LAVV repair evaluation. Although challenges remain in accurately modeling leaflet closure and chordal mechanics, this proof-of-concept work highlights the platform's potential for refining repair strategies before clinical application, which may be particularly relevant in small and heterogeneous populations with congenital heart disease.

## INTRODUCTION

Each year, 2100 children are born with an atrioventricular canal (AVC) in the United States [1]. This defect is even more prevalent among children with Down Syndrome—40% of those with cardiac disease have AVCs. In cases of complete atrioventricular canal (CAVC), rather than having distinct mitral and tricuspid valves, patients have a single common atrioventricular valve spanning both ventricles. Surgical repair aims to divide this single valve into two functional valves, including a mitral-equivalent left atrioventricular valve (LAVV) [2]. The resulting LAVV is comprised of two bridging leaflets that typically sutured together (closing the "cleft") to create a "neo-anterior" leaflet, as well as a mural leaflet.

Despite such corrective efforts, 5-20% of LAVV repairs will require reintervention, defined as subsequent surgical or catheter-based procedure following the initial CAVC repair. After the initial first intervention, an additional 37-40% of cases after the initial reintervention [3-6]. The repeat operations carry significant risks, with more than a quarter of patients requiring two or more reoperations, often associated with poor long-term outcomes and increased morbidity [5, 7]. The most common cause of reintervention is left atrioventricular valve regurgitation (LAVVR), primarily due to mechanical issues such as leaflet prolapse, leaflet tethering, ruptured chordae, residual clefts, annular dilatation, or dehiscence of the initial repair [5, 6, 8, 9]. Although moderate or greater LAVV regurgitation is the most common indication for reoperation, it is not the sole factor influencing long-term outcomes. While stenosis of the LAVV is rarely an indication for surgery on its own, studies have shown that the coexistence of a residual inflow gradient and mild or greater LAVVR significantly increases the risk of reintervention. In such cases, up to 73% of patients may require additional surgery, likely due to the added complexity in achieving a durable repair [6]. Other key predictors of reintervention include age at LAVV repair younger than 72 months, AVC anatomy without a ventricular septal defect, left ventricular dysfunction, and multiple regurgitant jets on postoperative transesophageal echocardiography (TEE) [10]. For pediatric patients, repairing a regurgitating valve is strongly preferred over replacement, as repairs accommodate patient growth and avoid complications like anticoagulation and multiple reoperations to replace mechanical valves to adjust for somatic growth associated with mechanical valves [11]. However, current repair techniques, including cleft closure, annuloplasty, commissuroplasty, leaflet augmentation, artificial chordae placement, and papillary muscle splitting, often fail to durably improve valve function [6].

Advancements in 3D echocardiography (3DE) have enabled detailed analyses of cardiac structures, providing precise, quantitative comparisons between normal and dysfunctional valves [12-14]. Leaflet tenting and billow, for example, are recognized predictors of valvular competence and surgical outcome, as they reflect the degree of leaflet displacement and leaflet redundancy, respectively. Assessing these parameters in vivo can help identify high-risk morphological features that may influence the success of specific repair strategies [13, 15]. However, these imaging tools only offer a snapshot in time and alone cannot fully predict the pathological functions needed to evaluate the effectiveness of repair strategies. Most existing 3D-printed valve models are static, and thus fail to simulate dynamic components such as pressure gradients, leaflet motion, and valve mechanics, which are critical for assessing surgical outcomes [16, 17].

Dynamic physical simulations have been successfully developed for mitral and tricuspid valves, providing platforms to test and refine repair techniques under physiologically relevant conditions [18, 19]. This study aims to develop a dynamic, image-derived physical simulation platform for LAVV repair, adapting these workflows to the unique challenges of pediatric LAVVs. By bridging the gap between static 3D modeling and real-world surgical conditions, this approach could offer a physiologically relevant tool to assess repair techniques, identify high-risk morphological features, and refine surgical strategies before clinical application. Such a platform has the potential to enhance preoperative planning, optimize repair durability, and ultimately improve outcomes for children undergoing valve repair.

## METHODS

### *Patient Selection and Image Acquisition*

Patients who had undergone AVC repair and high-quality 3DE imaging were identified from an institutional database. Approval for this study was obtained from the Institutional Review Board. Two cases of regurgitant LAVV were selected: a 21-year-old female with moderate regurgitation originating centrally from a residual cleft in the neo-anterior leaflet (Patient 1) and a 4-year-old male with severe regurgitation thought to be caused by dehiscence of a previous cleft closure (Patient 2). These existing 3DE images had been acquired via transesophageal echocardiography using a Philips EPIQ system (Philips Medical, Cambridge, MA). Selected patient information is outlined in **Table 1**, associate 3D echo and 3D color doppler echocardiography are shown in **Figure 1**.

### *Leaflet Segmentation and ValveMoldCreator*

Digital Imaging and Communications in Medicine (DICOM) files were imported into 3D Slicer (www.slicer.org). The annulus was defined in mid-systole and late-diastole using the *Annulus Analysis* module in the SlicerHeart extension [20]. The 3D Slicer platform is available at http://www.slicer.org, and open-source code and documentation for the *Valve Mold Creator* Module can be accessed at http://www.github.com/SlicerHeart. The mid-systolic frame is defined as the median time frame between the first frame when the valve is closed and the last frame when the valve is closed in a cardiac cycle. The late-diastolic frame occurs just before systole where the leaflets have not yet coapted and leaflet boundaries remain distinct. The leaflets in each phase were then manually segmented using the *Valve Segmentation* module and the papillary muscle tips were located to model two adjustable papillary muscles (**Figure 2a**). The identified papillary muscle tip locations were exported to Fusion 360 (Autodesk, San Rafael, CA) to design adjustable, patient-specific papillary muscle posts.

Late-diastolic leaflet molds were subsequently generated using custom code implemented in the *ValveMoldCreator* module in SlicerHeart (**Figure 2b**). This module extrudes a cylindrical mold from the atrial side of the leaflet segmentation, bounded by the user-defined annular curve and extended into the atrium. A virtual Hinge-Point Apparatus (HPA) was also created based on the annular curve (**Figure 1b**), enforcing anatomically accurate leaflet attachment at the annulus in the final valve design. Within *ValveMoldCreator*, various parameters can be adjusted, including mold height, leaflet coaptation boundaries, annulus enforcement, and HPA geometry, offering flexibility to replicate patient-specific anatomy and experimental variations (**Supplementary Video 1**). Finally, the valve profile mold, papillary muscle posts, and HPA were fabricated using an Ultimaker S3 3D printer (Utrecht, Netherlands), providing a physical framework for subsequent silicone valve construction.

### *Silicone Model Creation*

We created silicone models to emulate valve physiology that include leaflets, chordae, papillary muscles, and the surrounding myocardium. The leaflets were created by painting a composite of silicone and nonwoven gauze onto the 3D printed valve profile (Ecoflex, Smooth-On Inc., Macungie, PA). Gauze was embedded in the silicone to provide tensile strength and to prevent suture dehiscence under simulated hemodynamic loading. Six frayed Dacron threads modeled the chordae tendineae, running through the 3D-printed papillary muscle posts and attaching to adjustable tensioning keys (**Figure 2c**). The myocardium was modeled using a firmer silicone than the leaflets (Dragonskin, Smooth-On Inc., Macungie, PA) injected into a 3D printed mold containing a hinge-point-apparatus. Once the silicone had cured, the entire apparatus was removed from the mold and fixed between two 3D printed flanges.

### *Pulse Duplicator*

The assembled valve apparatus was then mounted in a two-chamber dynamic simulator (Archetype Biomedical Inc, London, Ontario, Canada) which can replicate patient heart rate, stroke volume, and pressures at the time of image acquisition (**Figure 1d**). Performance and regurgitation of the simulated valve were visualized using a Philips X8 transesophageal echocardiographic probe on an EPIQ echocardiographic machine (Philips Medical, Cambridge, MA) with EKG gating and color Doppler using ultrasound contrast (Lumason, Bracco Diagnostics INC, Monroe Township, NJ). 3D Zoom and Full Volume acquisitions were used with frame rates ranging from 28-45 frames per second.

### *Validation*

Echocardiographic data collected from the pulse duplicator were used to assess differences between the manufactured models and the ground-truth patient images. The image data were imported into 3D Slicer and the leaflets were segmented in mid-systole for analysis. The selected parameters for comparison are annular circumference, distance between commissures, annulus height, tenting height, and billow height. These metrics are previously defined and automatically measured using the *Valve Quantification* module in SlicerHeart [21].

Comparisons between five identical manufactured valves were performed to evaluate variance in the manufacturing process. The coefficient of variation (CV) was used to quantify variation in the annular and leaflet measurements. Variance was categorized as low (CV ≤ 10%), moderate (CV between 10% and 20%), or high (CV > 20%).

### *Comparison of Different Repair Strategies*

Five identical silicone valves for each patient were constructed and presented to an attending cardiothoracic surgeon to perform four different surgical valve repairs on four with the fifth used as a control. The strategies employed included: (1) annuloplasty where the annular circumference between the anterolateral commissure and the posteromedial commissure was reduced by 10%, (2) annuloplasty where the circumference was reduced by 20%, (3) Alfieri stitch between the neo-anterior leaflet and the mural leaflet, and (4) closure of the residual cleft in the neo-anterior leaflet.

Image data recorded from the pulse duplicator were imported into 3D Slicer, and the leaflets were segmented in mid-systole for analysis. The four different repair strategies were compared via common clinical metrics: billow height, tenting height, mean trans-valvar pressure gradient, and regurgitant orifice area. Billow height and tenting height are defined as the distance normal to the least-squares annular plane to the highest and lowest points on the leaflets, respectively [22]. The regurgitant orifice area was calculated using a custom shrink-wrapping algorithm on the resulting leaflet segmentation [23].

## RESULTS

A visual comparison of the mid-systolic segmentations derived from patient 3DE and corresponding segmentation of manufactured valves are shown in **Figure 3**. A quantitative assessment of valve geometry for model-generated segmentations and the corresponding patient valve segmentations is shown in **Table 2**. **Supplementary Video 2** provides an overview of the modeling workflow, including dynamic testing of the fabricated valve in the pulse duplicator, 3DE image acquisition and analysis, and demonstrations of simulated repair strategies.

Five identical unrepaired valve models were analyzed to assess manufacturing consistency. Annular metrics including annular circumference, commissural distance, and annulus height demonstrated low to moderate variability (CV = 2.1%, 4.1%, and 14.7% respectively). In contrast, leaflet closure metrics such as billow height, billow volume, tenting height, and tenting volume showed moderate to high variability (CV = 11.1%, 18.9%, 45.9%, and 73.5% respectively).

Comparisons of leaflet metrics across different repair strategies for Patient 1 and Patient 2 are presented in **Table 3**. For Patient 1, billow height and mean gradient pressure were largely consistent across all models, except for the Alfieri stitch repair, which exhibited a significantly elevated transvalvular mean pressure gradient indicative of a stenotic valve. Relative to the unrepaired valve model, tenting height increased in all repairs, reflecting mural leaflet immobility. Notably, both the Alfieri stitch and closure of the partial cleft in the neo-anterior leaflet eliminated any visible regurgitant area in the mid-systolic segmentations. It was also observed that after continuous use in the pulse duplicator, the sutured region of the valve repaired with an Alfieri stitch began to degrade and separate indicating that tension in the leaflet edges and sutures induced by the hemodynamic pressure was too great for the implementation of this repair. In Patient 2's simulated repairs, mean pressure gradient was elevated in all strategies; although no single approach completely eliminated regurgitation, the introduction of an 11 mm patch augmentation to bridge the existing cleft and the use of a commissuroplasty produced the greatest reduction in regurgitant area.

## DISCUSSION

The LAVV is a complex surgically created valve whose dynamic behavior is influenced by multiple factors, including leaflet geometry, subvalvar apparatus, and hemodynamic conditions. Capturing these factors in a physical simulation is critical for understanding valvular pathology and evaluating real-time responses to surgical interventions—particularly in congenital atrioventricular canal (AVC) defects, where anatomy varies significantly between patients. Unlike adult mitral valve disease, which often follows more reproducible patterns, congenital heart disease cases rarely present two alike. This makes it difficult for surgeons to generalize techniques across cases, and opportunities for gaining hands-on experience are limited. Patient-specific models thus provide a critical platform for preoperative rehearsal, especially in settings where standardization is not possible. This study demonstrates the feasibility of a dynamic, image-derived modeling platform for LAVV repair but also highlights challenges and areas for refinement.

Pronounced discrepancies arose when the manufactured valves were compared to ground-truth clinical echocardiograms, with only annular circumference and annular height in Patient 1 showing relatively small errors (2.1% and 4.2% respectively). Most parameters—particularly those describing leaflet dynamics—did not accurately reproduce the in vivo physiology. The most likely explanation is the passive nature of the current experimental setup, wherein valve closure is driven solely by pressures in the pulse duplicator. In a clinical scenario, LAVV closure is an active process involving coordinated myocardial contraction, papillary muscle traction, and annular displacement. Integrating these active forces into the simulation could potentially reduce discrepancies in leaflet motion and improve coaptation fidelity.

The analysis of five identical, unrepaired valve models highlights distinct variability patterns in annular and leaflet closure metrics. Annular metrics, including annular circumference, commissural distance, and annulus height, exhibited low to moderate variability. This consistency suggests that the annular shape is highly preserved through the *ValveMoldCreator* module and manufacturing process. Conversely, leaflet closure metrics showed moderate to high variability, reflecting the increased complexity of leaflet closure dynamics in the models. This variability can be attributed to inconsistencies in silicone leaflet thickness and chordal tensions, which propagate into larger variations in leaflet shape and position during systole. Furthermore, only primary chordae, which extend from leaflet edges to the papillary muscles, are simulated in these models. Secondary and tertiary chordae, which connect the leaflet body to the ventricular wall and help relieve tension within the leaflet tissue, are not included in this simulation [24]. These omissions likely contribute to the observed discrepancies, underscoring the challenges of fully replicating the intricate mechanics of leaflet dynamics in physical models.

Lastly, in the comparison of different repair strategies in the two selected patients, we were able to completely eliminate regurgitation in Patient 1 through closing the residual cleft and by using and Alfieri stitch. However, after long runs in the pulse duplicator, the Alfieri stitch began to stretch and degrade. This created a central orifice through which leaking was observed, rendering the repair ineffective over time. Additionally, the creation of a double orifice introduced mild stenosis, as the valve opening became partially obstructed. The two annuloplasty repairs left residual regurgitation but resulted in much lower mean pressure gradients that avoid stenosis. These findings suggest that while mildly restricting flow across the valve, the cleft closure would provide the most relief as a reintervention.

For Patient 2, no definitive optimal repair strategy was identified. Lack of a control model or a true clinical ground truth precluded definitive comparisons, and all simulated repairs yielded elevated mean pressure gradients indicative of varying degrees of stenosis. Manufacturing inconsistencies, particularly regarding silicone thickness in the leaflets, further complicate interpretation. Additionally, simplifications of the valvar and sub-valvar apparatuses reduce physiological accuracy. Addressing these limitations will be crucial for the continued refinement of LAVV physical models and for drawing more robust conclusions regarding optimal repair strategies.

The chief limitations of this work are the passive nature of the physical simulation, and difficulties of reliably replicating realistic material properties. Future iterations of this platform should incorporate active myocardial and annular motion and more sophisticated chordal structures to capture in vivo valve mechanics more accurately. Improvements in material science, such as 3D-printed tissue-mimicking materials with variable stiffness, may enhance the consistency of leaflet thickness and mechanical properties. Inclusion of advanced sub-valvar apparatuses, such as replicated secondary and tertiary chordae, would help distribute stresses more realistically throughout the leaflets. However, while improvement is possible, the labor-intensive process and difficulty in reproducing material properties is likely prohibitive to practical application of image-derived physical models of LAVV for clinical planning. While physical models enable the opportunity for surgeons to physically practice repairs, emerging *in silico* approaches to iterative pre-operative optimization of repair may provide a more practical long term solution, but also require further refinement and validation [23, 25, 26].

## CONCLUSION

Overall, this work demonstrates that a dynamic, image-derived physical simulation can replicate key anatomical features and facilitate the preclinical assessment of LAVV repair strategies but is fundamentally limited by the difficulty in realistically replicating the leaflet material properties and biomechanics. By addressing the identified limitations and incorporating more physiologically accurate features, the simulation has the potential to inform surgical planning and minimize reinterventions. However, emerging in silico-based modeling may be more practical and realistic for clinical application.

**Funding:** Topolewski Endowed Chair in Pediatric Cardiology, the Topolewski Pediatric Valve Center (Frontier Program), and NIH R01HL153166.

**Statements and Declarations:** Elvis Chen, John Moore, and Terry Peters are co-founders of Archetype Biomedical. The other authors have no disclosures.

## Figures Legends

**Figure 1: 3D Echocardiograms of Patients Used for Modeling:** 3D echo volume renderings and 3D color doppler echocardiography of the two selected patients. Patient 1 (top) shows a regurgitant jet originating from the mid portion of the anterior (fused superior and inferior bridging leaflet) and the mural (posterior leaflet) cooptation region. Patient 2 (bottom) shows a larger jet originating from the cleft of the anterior (fused superior and inferior bridging leaflet).

**Figure 2: Overview of Model Creation.** A. The 3D echo is segmented and the annulus is defined using SlicerHeart; B) The ValveMoldCreator is then used to construct the a mold of the atrial surface of the leaflet segmentation as well as the HPA; C. The mold and HPA are then 3D printed before being used to fabricate a physical valve model; D. The assembled valve model with simulated sub-valvar apparatus are inserted into the pulse duplicator and function under physiological conditions.

**Figure 3: Comparison of Models of Original Image and Model from Echocardiogram of Resulting Physical Model.** The mid-systolic segmentations derived from the original patient imaging (left) and the imaging of the fabricated valves acquired from the pulse duplicator (right). The neo-anterior leaflet is segmented in red, and the left mural leaflet is segmented in orange. Mid-systolic annuli are contoured and shown as a red curve.

**Figure 4: Comparison of Model to 3D Echocardiographic Images of Model.** Atrial view of each repair strategy tested (top) and 3DE images of each repaired valve for Patient 1 in the pulse duplicator (bottom).

***Figure 5:*** **Comparison of Model to 3D Echocardiographic Images of Model.** Atrial view of each repair strategy tested (top) and 3DE images of each repaired valve for Patient 2 in the pulse duplicator (bottom).

**Figures (For Review)**

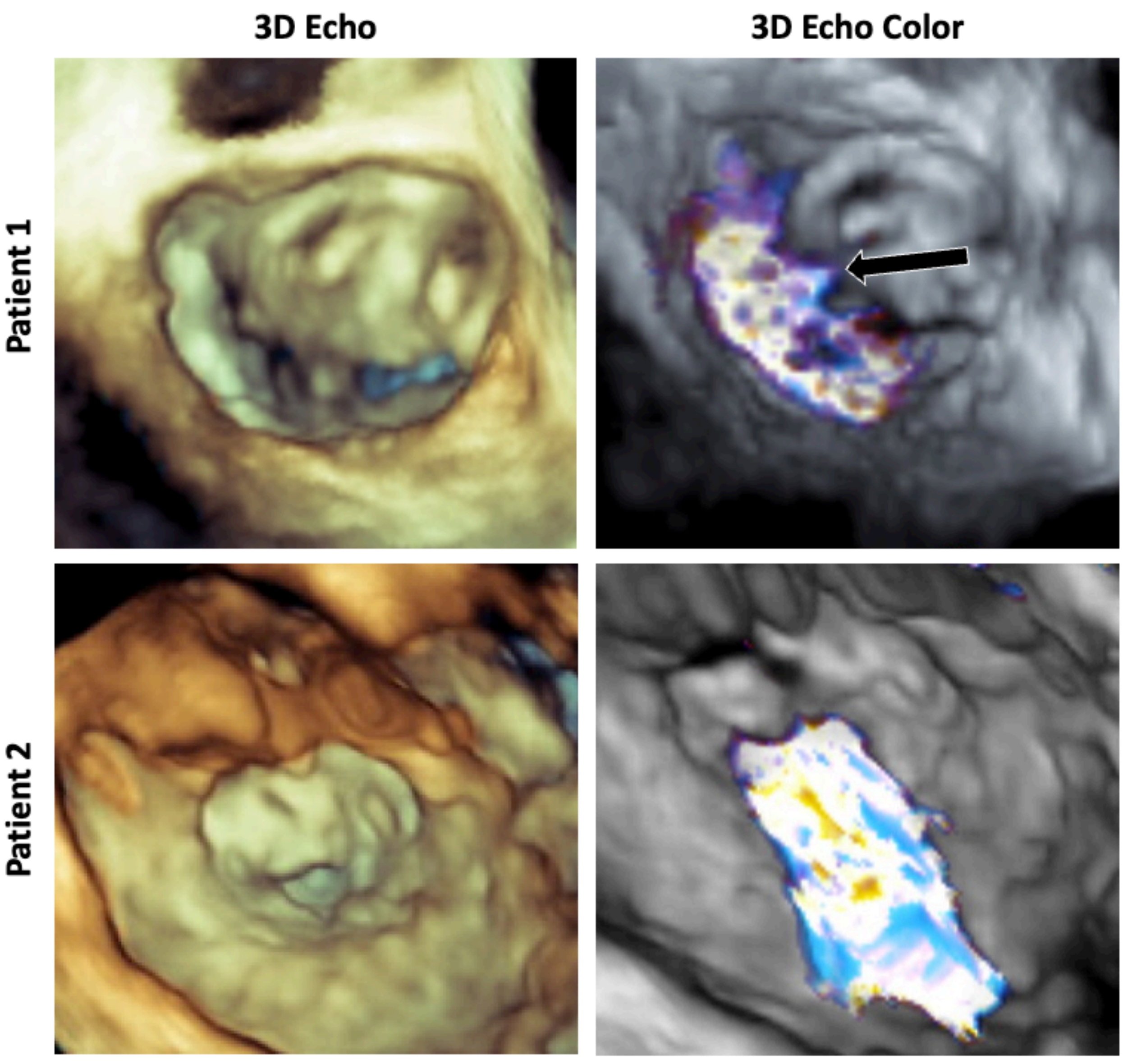

**Figure 6: 3D Echocardiograms of Patients Used for Modeling:** 3D echo volume renderings and 3D color doppler echocardiography of the two selected patients. Patient 1 (top) shows a regurgitant jet originating from the mid portion of the anterior (fused superior and inferior bridging leaflet) and the mural (posterior leaflet) cooptation region. Patient 2 (bottom) shows a larger jet originating from the cleft of the anterior (fused superior and inferior bridging leaflet).

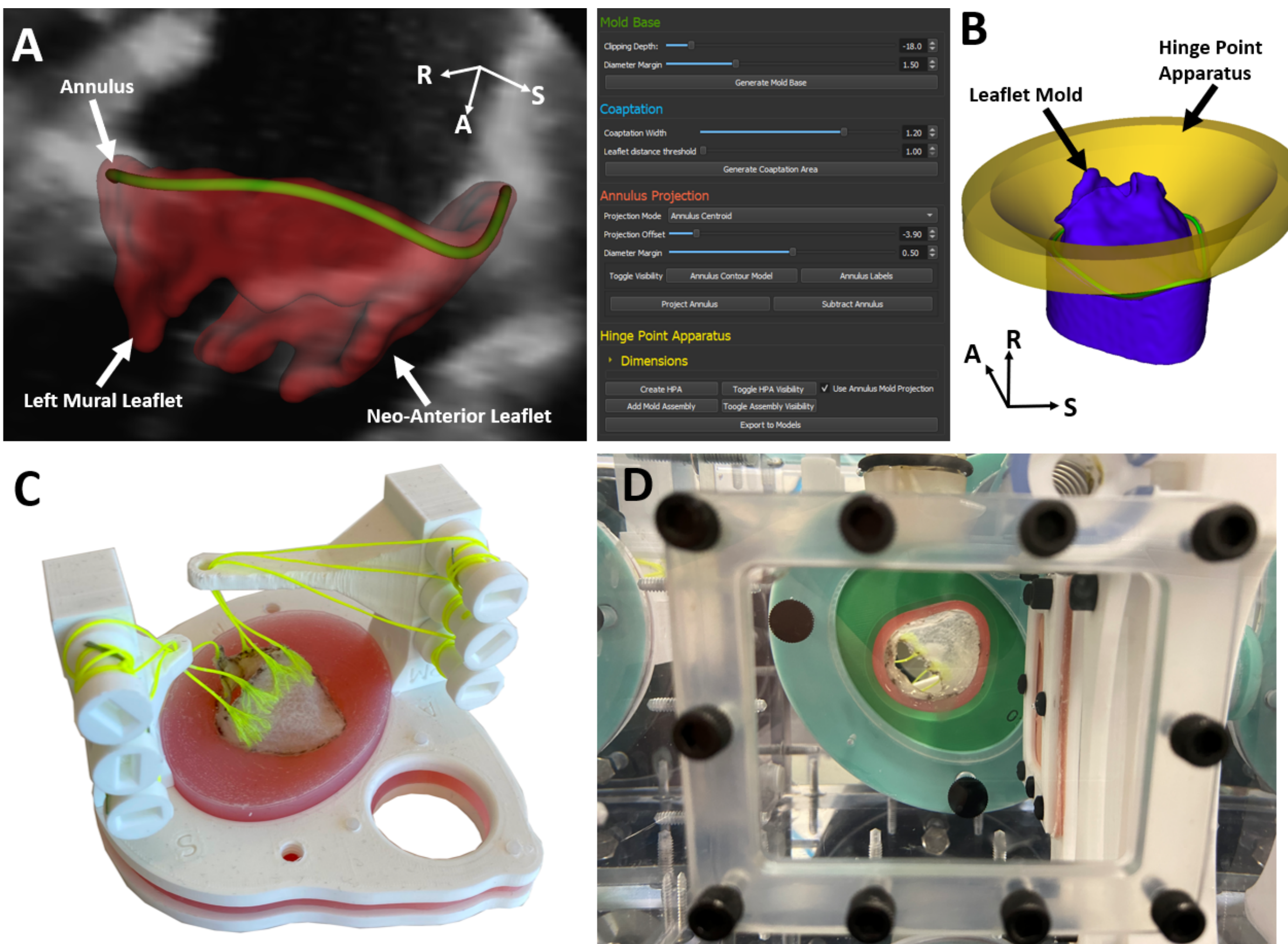

**Figure 7: Overview of Model Creation.** A. The 3D echo is segmented and the annulus is defined using SlicerHeart; B) The ValveMoldCreator is then used to construct a mold of the atrial surface of the leaflet segmentation as well as the HPA; C. The mold and HPA are then 3D printed before being used to fabricate a physical valve model; D. The assembled valve model with simulated sub-valvar apparatus is inserted into the pulse duplicator and function under physiological conditions.

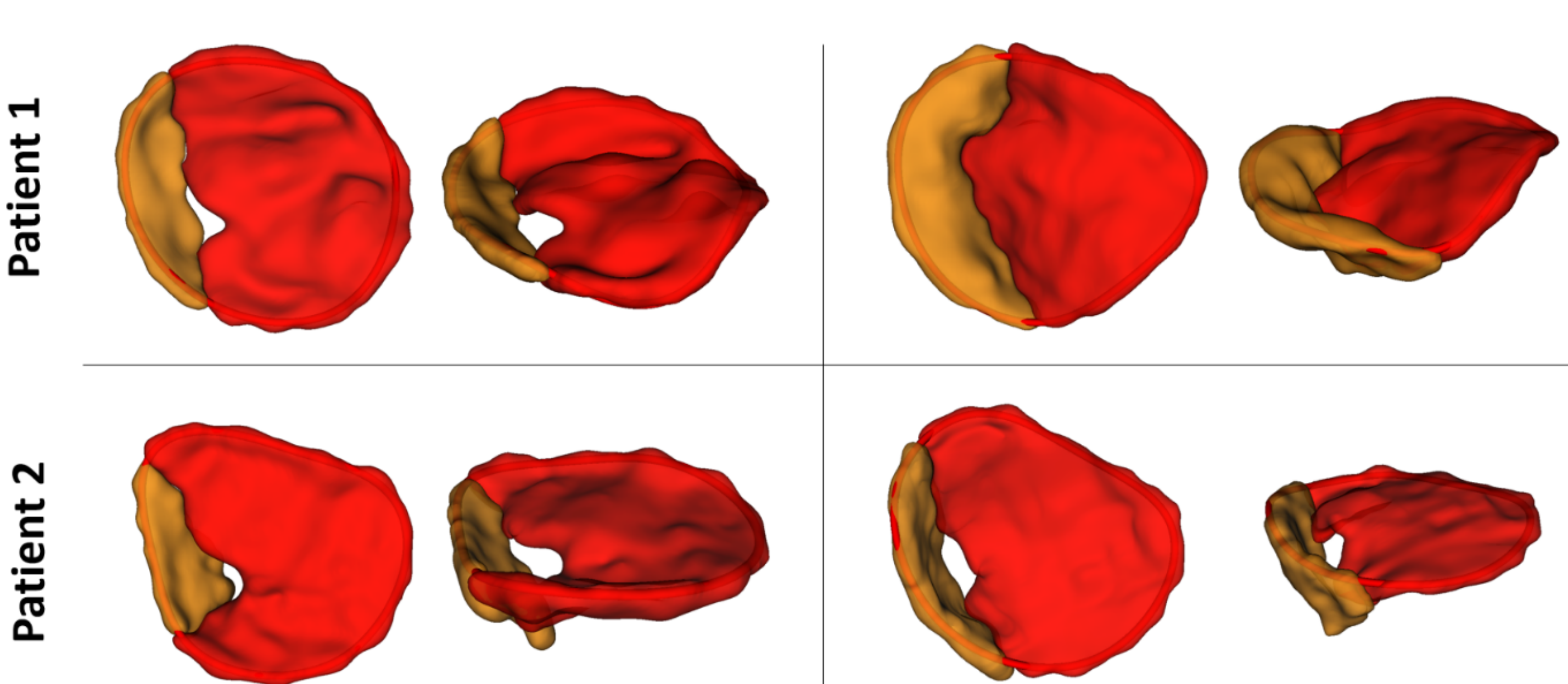

**Figure 8: Comparison of Models of Original Image and Model from Echocardiogram of Resulting Physical Model.** The mid-systolic segmentations derived from the original patient imaging (left) and the imaging of the fabricated valves acquired from the pulse duplicator (right). The neo-anterior leaflet is segmented in red, and the left mural leaflet is segmented in orange. Mid-systolic annuli are contoured and shown as a red curve.

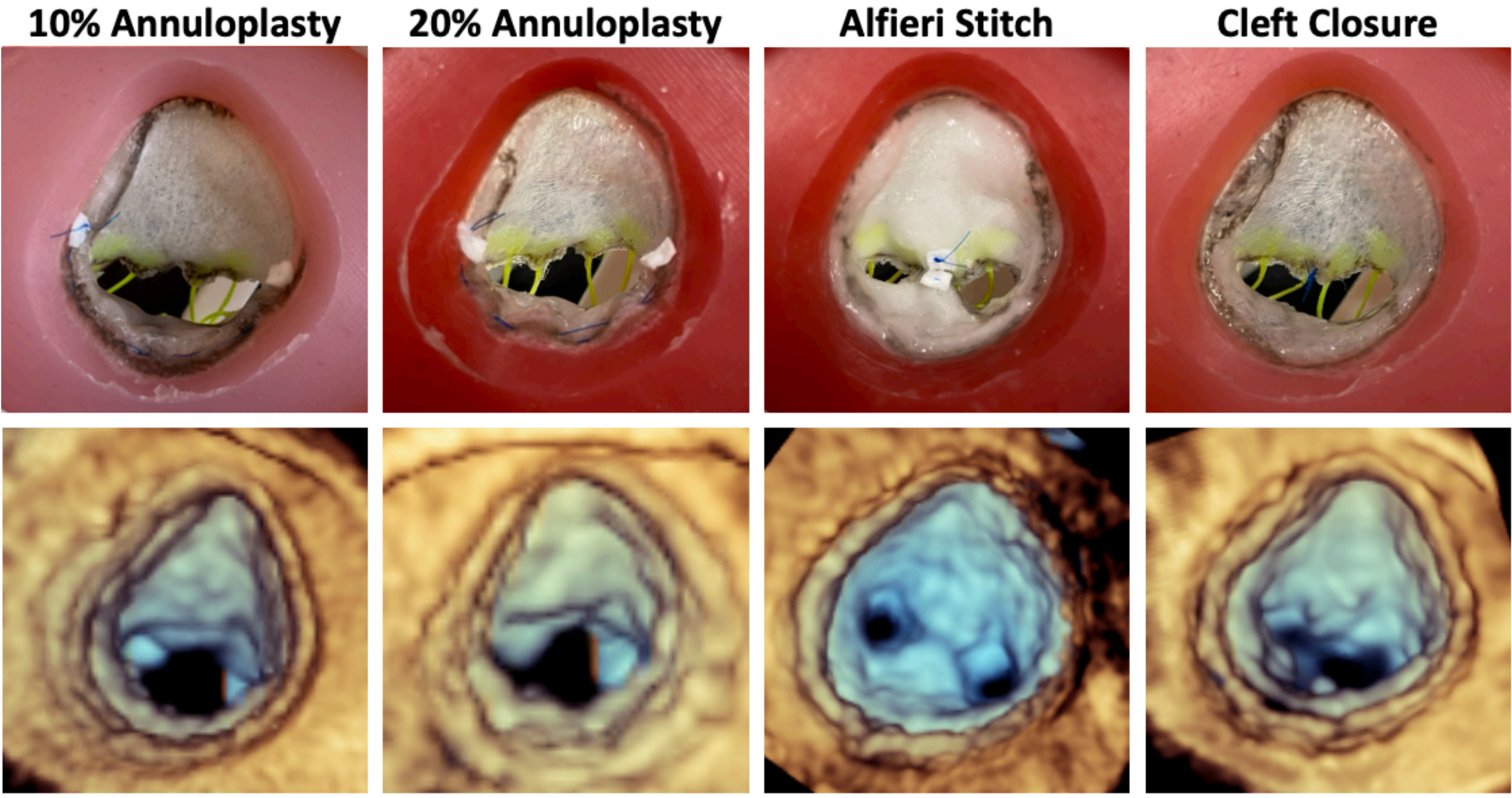

**Figure 9: Comparison of Model to 3D Echocardiographic Images of Model.** Atrial view of each repair strategy tested (top) and 3DE images of each repaired valve for Patient 1 in the pulse duplicator (bottom).

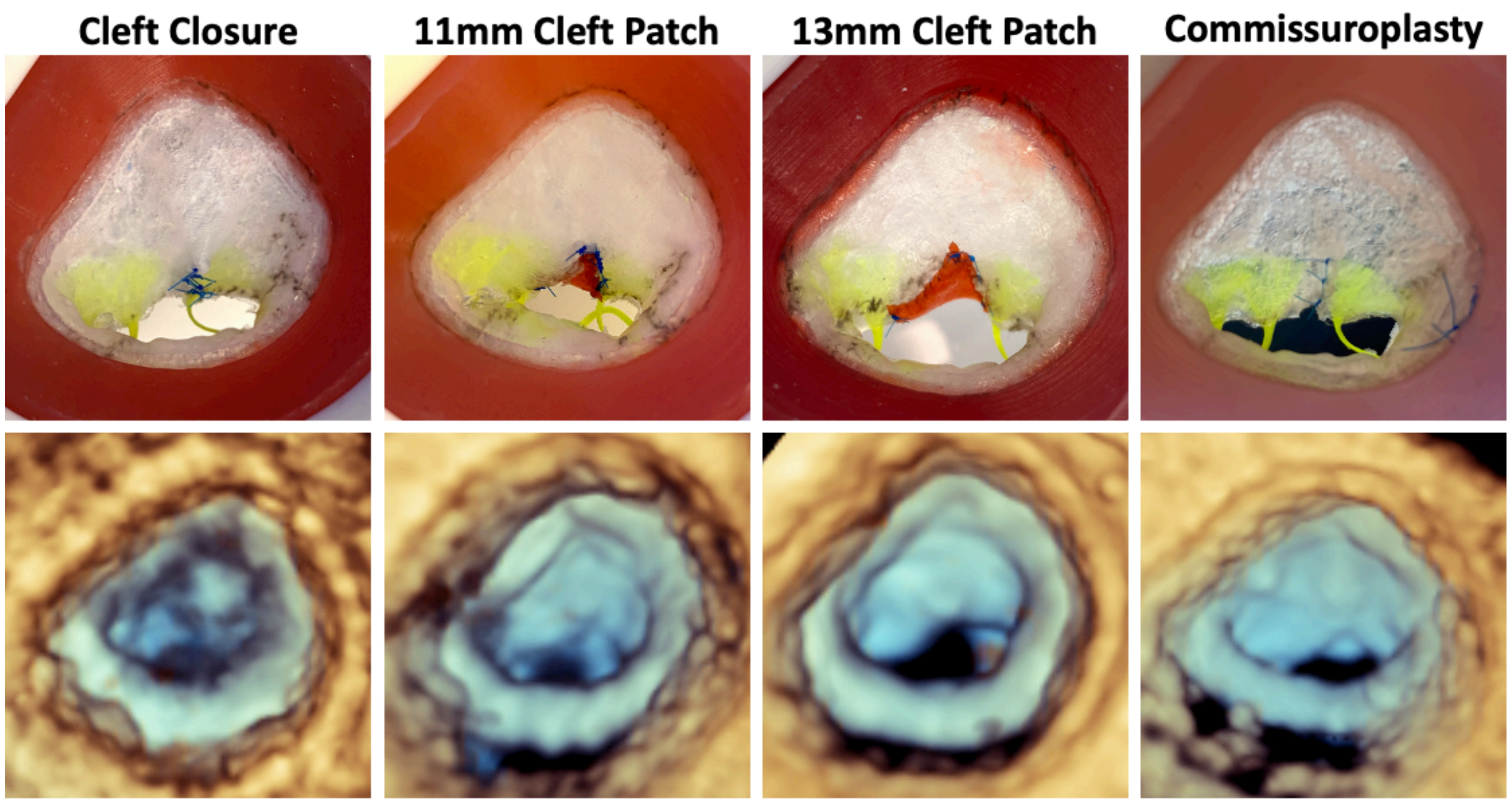

***Figure 10:*** **Comparison of Model to 3D Echocardiographic Images of Model.** Atrial view of each repair strategy tested (top) and 3DE images of each repaired valve for Patient 2 in the pulse duplicator (bottom).

## Tables

*Table 1. Patient information*

| | Age (years) | Weight (kg) | BSA ($m^2$) | Diagnosis | Surgical Repair on LAVV |
|---|---|---|---|---|---|
| **Patient 1** | 21 | 45.0 | 1.12 | Incomplete AV canal s/p repair, closed cleft, moderate regurgitation originating from A2-P2 coaptation defect | Cleft Closure; 2 sutures placed at apex of neo-anterior leaflet |
| **Patient 2** | 4 | 12.7 | 0.57 | Transitional AVC, partial cleft, eccentric bridging leaflets (superior much larger than inferior), regurgitant jet originating from cleft dehiscence | Attempted cleft closure resulting in residual regurgitation. Replaced with mechanical value |

*Table 2. Comparison of clinical and manufactured model metrics*

| Variable | Patient 1 | | | Patient 2 | | |
|---|---|---|---|---|---|---|
| | **Clinical Image** | **Model Image** | **% Difference** | **Clinical Image** | **Model Image** | **% Difference** |
| **Annulus Circumference (cm)** | 11.2 | 11.6 | 3% | 10.2 | 10.7 | 5% |
| **Commissural Distance (cm)** | 2.9 | 3.5 | 19% | 2.2 | 3.3 | 41% |
| **Annulus Height (cm)** | 0.5 | 0.5 | 0% | 0.4 | 0.3 | 28% |
| **Billow Height (cm)** | 0.4 | 0.3 | 28% | 0.2 | 0.5 | 85% |
| **Tenting Height (cm)** | 0.4 | 0.7 | 54% | 0.6 | 0.4 | 40% |
| **Billow Volume ($cm^3$)** | 1.0 | 0.9 | 13% | 0.5 | 1.7 | 109% |
| **Tenting Volume ($cm^3$)** | 0.5 | 0.7 | 33% | 0.4 | 0.2 | 67% |

*Table 3. Comparison of repair strategies*

| | Repair Type | Billow Height (cm) | Tenting Height (cm) | Mean Gradient (mmHg) | Regurgitant Area ($cm^2$) |
|---|---|---|---|---|---|
| **Patient 1** | 10% Annuloplasty | 0.30 | 1.05 | 4 | 0.991 |
| | 20% Annuloplasty | 0.30 | 0.89 | 5 | 1.093 |
| | Alfieri Stitch | 0.35 | 0.76 | 17 | 0 |
| | Cleft Closure | 0.34 | 0.90 | 9 | 0 |
| **Patient 2** | Direct Cleft Closure | 0.54 | 0.62 | 11 | 0.274 |
| | 11mm Patch Augmentation | 0.62 | 0.37 | 7 | 0.147 |
| | 13mm Patch Augmentation | 0.52 | 0.40 | 10 | 0.258 |
| | Commissuroplasty | 0.58 | 0.39 | 13 | 0.149 |

**Supplementary Videos:**

**Video 1: Demonstration of Valve Mold Creator Module**
**Video 2: Overview dynamic testing of the fabricated valve in the pulse duplicator, 3DE image acquisition and analysis, and demonstrations of simulated repair strategies**